# First principles investigation of structural, electronic and optical properties of MgRh intermetallic compound

**Md. Lokman Ali[1], Md. Zahidur Rahaman[2], Md. Atikur Rahman*[3], Md. Afjalur Rahman[4]**

[1, 2, 3,4]Department of Physics, Pabna University of Science and Technology, Pabna-6600, Bangladesh

**Abstract**

The structural electronic and optical properties of intermetallic compound MgRh were investigated by using the ab-initio technique from CASTEP code. In this study we have carried out the pseudo-potential plane-wave (PP-PW) method based on the density functional theory (DFT), within the generalized gradient approximation (GGA). Our calculated structural parameters and corresponding graphical values fit with other previous available experimental data and other theoretical observations. The calculated electronic band structure reveals metallic conductivity and the major contribution comes from Rh-d states. Comparison between our investigated properties and experimental data shows good agreement. The optical functions (dielectric functions, refractive index, absorption spectrum, conductivity, energy loss spectrum and reflectivity) have been calculated and discussed. This is the first quantitative prediction of the electronic and optical properties of intermetallic compound MgRh alloy, since it has not been reported yet. The calculated optical functions reveal that the reflectivity is high in the ultraviolet region up to 73 eV for MgRh, showing this to be promising coating materials.

## 1. Introduction

Magnesium (Mg) is one of the important alloy and most abundant element in the Earth. Magnesium has always been great attractive strong interest among the research community due to their remarkable physical properties. The researches on Magnesium alloys are to explore due to its low density (approximately 1.74 gm/cm$^3$) and high specific toughness and stiffness than many other alloys including aluminium and have bigger tensile strength than steel. Magnesium and its alloys also shows a number of different interesting properties such as a high specific strength, good ductility, high damping capacity, good machinability, high thermal conductivity, electromagnetic shielding, good corrosion resistance and high recycling potential [1]. Magnesium alloy is one of the lightest materials, namely as the "green" engineering materials [2, 3]. The element and its alloys play an important role of modern industry needs. These alloys have been used in a wide variety of prominent applications, especially in automotive industry, national defense industry and aerospace manufacturing [4]. The above remarkable properties and prominent application of Mg motivated us to study of this alloy.

First principles calculation or ab initio methods is a new state of condensed matter Physics and so we have used it to calculate the structural, electronic and optical properties of MgRh compound for the first time. Most of the theoretical and experimental works have been carried out on the elastic, mechanical and thermodynamic properties of intermetallic compounds MgRh. V.B. Compton [5] studied the crystal structure of MgRh compound by using X-ray powder diffraction method. The elastic, mechanical and thermodynamic properties of this compound have been investigated by S. Boucetta [6] theoretically. Pingli Mao et al. [7] have studied the structural, elastic and electronic properties of AB$_2$ type intermetallics in Mg-Zn-Ca-Cu alloy.

*Corresponding Author: atik0707phy@gmail.com

H. Takamura et al. [8] have also investigated the novel compound of MgRh with a CsCl-type structure. To the best of our knowledge, the electronic and optical properties of MgRh compounds with CsCl-type structure are still unexplored. It is well known that an optical property provides very useful information for calculating electronic band structure and magnetic excitation. The main objective of this paper is to investigate the structural, electronic and optical properties of MgRh intermetallic compounds for the first time by using First Principle calculations (or ab initio method). The presentation format of this paper is organized as follows: section 2 briefly describes the computational method. Section 3 presents the results of the analysis and discussion. Finally, section 4 presents summary of our results and conclusions.

## 2. Computational details

The density functional theory (DFT) based CASTEP computer program with the generalized gradient approximation (GGA) with the PBE exchange correlation function [9-13] is used to perform the calculations. Mg-$2P^6$ $3S^2$ and Rh-$4d^8$ $5S^1$ are treated as valence electrons of MgRh. The electromagnetic wave functions were expanded in a plane wave basis set. We have used energy cut-off of 340 eV to perform the calculations. The Monkhorst-Pack scheme [14] is used to construct the K-point sampling of Brillouin zone. We have used 4×4×4 grids in primitive cells of MgRh. The equilibrium crystal structures were obtained via geometry optimization in the Broyden-Fletcher-Goldfarb-Shanno (BFGS) minimization scheme [15]. The criteria of convergence in the geometry optimization were set to $1.0 \times 10^{-5}$ eV/atom for energy, $1 \times 10^{-3}$ Å for ionic displacement, 0.03 eV/Å for force, and 0.05 GPa for stress. These parameters were carefully verified to lead a better converged total energy.

## 3. Results and discussion

### 3.1. Structural properties

The MgRh intermetallic compound belongs to cubic crystal structure. It has CsCl-type structure with space group Pm-3m (221). The value of equilibrium lattice parameter is 3.099 Å [5]. By minimizing the total energy the structure has been optimized as a function of normal pressure. The optimized crystal structure of MgRh is shown in Fig.1. In Table 1 the calculated values of the structural properties of MgRh have been shown along with the available experimental and theoretical values. It is evident from Table 1 that our present theoretical results match well to both experimental and other theoretical results.

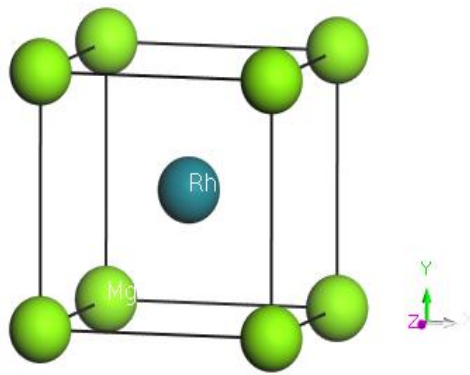

**Fig. 1.** The crystal structure of MgRh

**Table 1.** The calculated equilibrium Lattice constant "$a_0$", unit cell volume "$V_0$" and bulk modulus "$B_0$" of MgRh.

| Properties | Expt.[5] | Other Calculation[6] | Present Calculation | Deviation from Expt. (%) |
| --- | --- | --- | --- | --- |
| $a_0$ (Å) | 3.099 | 3.101 | 3.18 | 2.61 |
| $V_0$ (Å$^3$) | - | 29.791 | 32.15 | - |
| $B_0$ (GPa) | - | - | 129.41 | - |

From Table 1 one can see that the present calculated value of lattice constant is 3.18 Å which exhibits 2.61 % deviation from the experimental value. Our calculated value is slightly different from the other theoretical value due to different calculation methods. The experimental data are measured at room temperature while our investigated data are simulated at 0 K. Perhaps this is the reason for the existing discrepancy between the present theoretical value and experimental value. Hence the above discussion indicates the reliability of our present first principle calculations of intermetallic MgRh.

*3.2. Electronic properties*

The study of the density of states is crucial in the analysis of the physical properties of MgRh intermetallic. In order to know the details bonding nature of MgRh compound, we have calculated the the total density of states (TDOS) and partial density of states (PDOS) of MgRh as shown in Fig.3. To know about the metallic nature of MgRh the electronic band structures along the high symmetry directions in the Brillouin zones have been calculated and is shown in Fig.2.

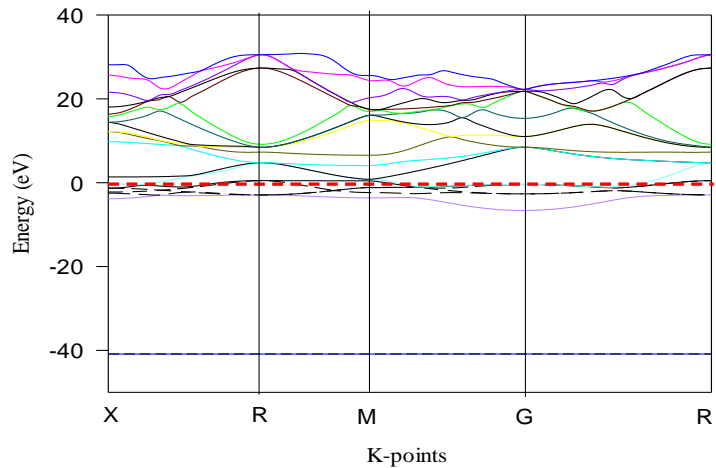

**Fig. 2.** Electronic band structure of MgRh along high symmetry direction in the Brillouin zones.

From the band structure diagram as shown in Fig.2, it is noticed that MgRh exhibits metallic nature since a number of conduction and valence bands are overlapping at the Fermi level. By analyzing the diagram of PDOS and TDOS as shown in Fig.3, we notice that the main bonding peaks locate in the energy range between 0 and -5 eV. The contribution of Mg-p and Rh-d states are dominant. However Rh-d orbital contributes largely than Mg-p orbital at the Fermi level.

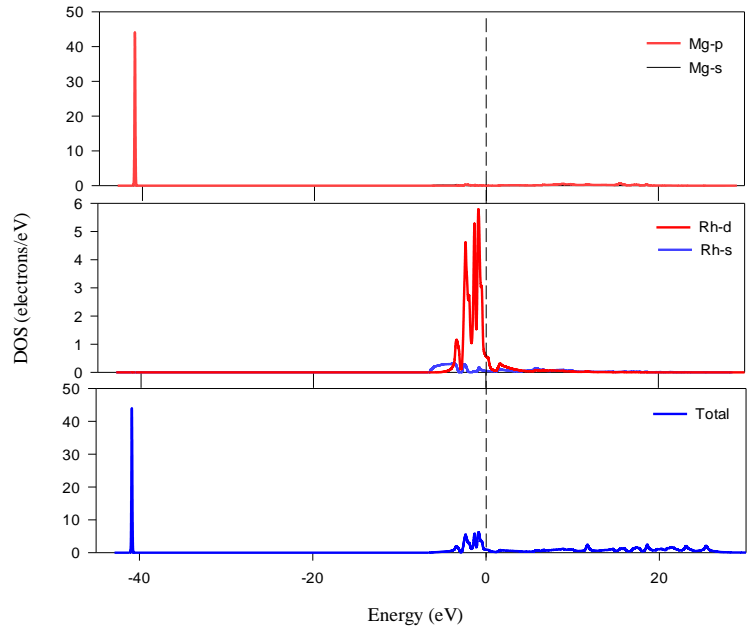

**Fig. 3.** The partial and total density of states of MgRh

The bonding electronic orbits determine the structural stability of intermetallic compound [16]. The formation of covalent bond in MgRh intermetallic compound is very weak since the ionic bond depends on the charge transfer between atoms, while covalent bond is determined by the depth and width of band gap near Fermi level.

For investigating the bonding nature of materials, Mulliken overlap population [17] is a great quantitative criterion. The calculated data of the atomic Mulliken population of MgRh compound is tabulated in Table 2. From Table 2, we see that the bond population is -1.99 which denotes the ionic nature of MgRh compound since the low value of bond population denotes the ionic nature of the materials [18].

**Table 2.** Mulliken electronic populations of MgRh.

| Species | s | p | d | Total | Charge | Bond | Population | Lengths |
|---|---|---|---|---|---|---|---|---|
| Mg | 0.69 | 5.90 | 0.00 | 6.58 | 1.42 | Mg-Rh | -1.99 | 2.755 |
| Rh | 1.13 | 0.83 | 8.45 | 10.42 | -1.42 | | | |

*3.3. Optical properties*

The frequency dependent dielectric function $\varepsilon(\omega) = \varepsilon_1(\omega) + i\varepsilon_2(\omega)$ is used to determine the optical properties of MgRh. Where $\varepsilon_2(\omega)$ is the imaginary part which is obtained from the momentum matrix elements between the filled and the unfilled electronic states and calculated using the below equation [19] -

$$\varepsilon_2(\omega) = \frac{2e^2\pi}{\Omega\varepsilon_0} \sum_{k,v,c} |\psi_k^c| u.r |\psi_k^v|^2 \delta(E_k^c - E_k^v - E) \qquad (1)$$

Where, *u* is defined as the polarization of the incident electric field, *ω* is defined as the frequency of light, *e* is the charge of electron, $\psi_k^c$ is the conduction band wave function and $\psi_k^v$ is the valence band wave function at *K* respectively. The Kramers-Kronig transform is used to derive the real part. From Eqs. 49 to 54 in ref. [19] are used to determine the all other optical constants. Fig.4 represents the optical functions of MgRh investigated for photon energies up to 80 eV. 0.5 eV Gaussain smearing is used for all calculations.

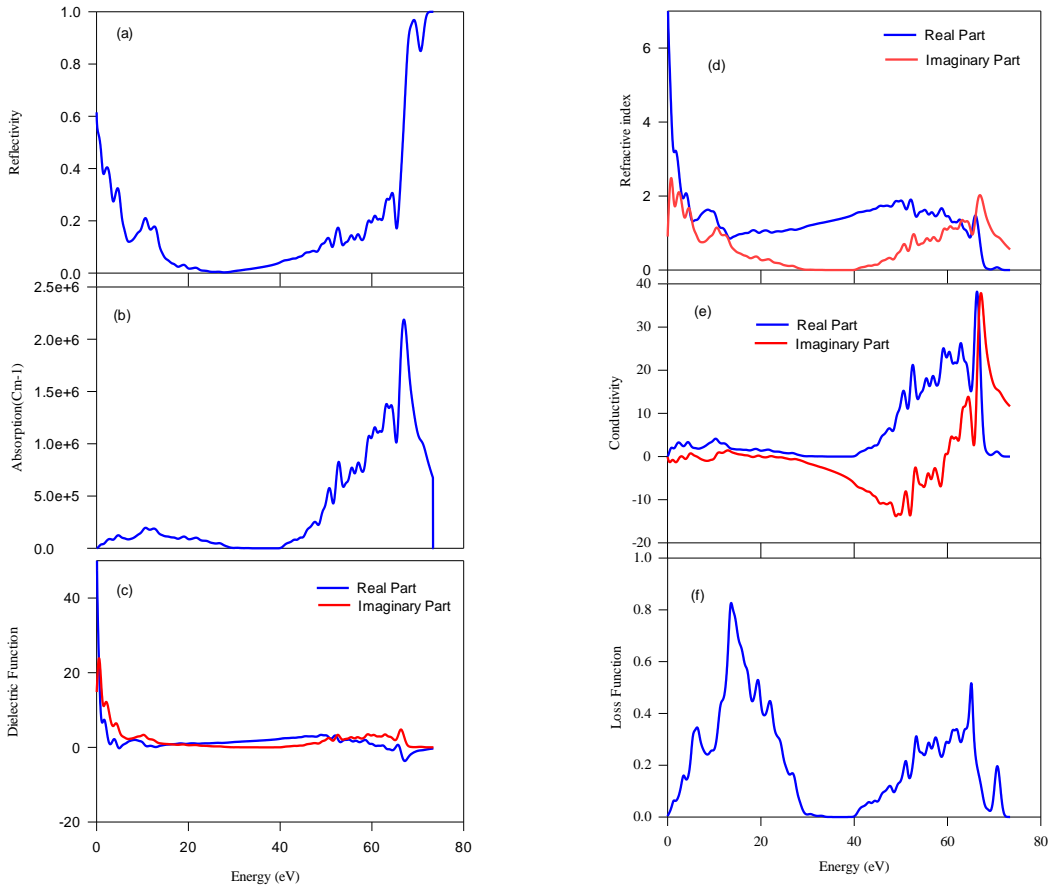

**Fig. 4.** The optical functions (a) reflectivity, (b) absorption, (c) dielectric function, (d) refractive index, (e) conductivity, and (f) loss function of MgRh intermetallic compound.

Fig.4 (a) represents the reflectivity spectra of MgRh. From this figure we see that the reflectivity is 0.50-0.68 in the infrared region. The value of the reflectivity drops in the visible region and increases in the ultraviolet region up to 73 eV. The absorption coefficient of MgRh is shown in Fig.4 (b), from which we see that the highest peak located at 65.25 eV. This value indicates that MgRh has good absorption coefficient in the ultraviolet region. The dielectric function of MgRh is represented in Fig.4 (c) as a function of photon energy. From this figure it can be seen that the imaginary part of dielectric function becomes zero at about 73 eV which indicates that above this energy MgRh becomes transparent. The region of nonzero value in the energy range from 0-12.50 eV, and 28-55.05 eV indicates the occurrence of absorption in that energy range. The static dielectric constant of MgRh intermetallic is 50. Fig.4 (d) shows the diagram of refractive index of MgRh. The value of static refractive index is 8, which decreases gradually in the visible and ultraviolet region. Fig.4 (e) illustrates the conductivity spectra of MgRh. From this figure we see that photoconductivity starts with zero photon energy, which indicates the metallic nature of MgRh. This result is similar with result getting from the band structure of MgRh shown in Fig.2. The energy loss function is an important optical parameter to describe the energy loss of a fast electron traversing the material and is large at the plasma frequency [20]. Fig.4 (f) represents the energy loss spectrum of MgRh as a function of photon energy. We found the prominent peak at about 17 eV, at which the reflectance is rapidly diminished.

## 4. Conclusions

In this paper, we have studied the structural, electronic and optical properties of intermetallic compound MgRh by using the generalized gradient approximation (GGA). Since the GGA-PBE method shows good agreement with LDA method [21], hence we have just run our investigation using GGA-PBE. The calculated lattice parameters show good agreement with the experimental and other theoretical values. The study of electronic properties shows that the compound under study is metallic in nature. The investigation of DOS shows that the major contribution near the Fermi level comes from Rh-d states. Further investigation on the Mulliken overlap population indicates that the ionic nature is dominant in MgRh compound. The study on optical properties of MgRh exhibits that the reflectivity is high in the ultraviolet region up to 73 eV. The absorption coefficient of MgRh is good in the ultraviolet region. We are unable to compare our investigated result since there is no previous result available for the electronic and optical properties of intermetallic MgRh. Just one literature is available for the investigation of the electronic and optical properties of MgCu at ambient condition [22]. Our investigated data is nearly similar with the calculated data of MgCu since both compounds are similar in nature. However we expect that this present study will motivate to experimental investigation on electronic and optical properties of intermetallic MgRh.